\documentclass[aps,preprint,amsmath,amssymb,showpacs,amsfonts,prd,nofootinbib]{revtex4-1}
\usepackage{epsfig}
\usepackage{graphicx}
\usepackage{dcolumn}
\usepackage{bm}
\usepackage{amsmath, amsthm, amssymb}
\usepackage{color}

\newcommand{\del}{\partial}

\newcommand{\bea}{\begin{eqnarray}}
\newcommand{\eea}{\end{eqnarray}}

\begin{document}
\title{A Novel Formula for Bulk Viscosity from the Null Horizon Focusing Equation}
\author{Christopher Eling$^1$}
\author{Yaron Oz$^2$}
\affiliation{$^1$ SISSA, Via Bonomea 265, 34136 Trieste, Italy and INFN Sezione di Trieste, Via Valerio 2, 34127 Trieste, Italy}
\affiliation{$^2$ Raymond and Beverly Sackler School of
Physics and Astronomy, Tel-Aviv University, Tel-Aviv 69978, Israel}
\date{\today}
\begin{abstract}
The null horizon focusing equation is equivalent via the fluid/gravity correspondence  to the entropy balance law of the fluid. Using this equation we derive a simple novel formula for the bulk viscosity of the fluid. The formula is expressed in terms of the dependence of scalar fields at the horizon on thermodynamic variables such as the entropy and charge densities. We apply the formula to three classes of gauge theory plasmas: non-conformal branes, perturbations of the ${\cal N}=4$ supersymmetric Yang-Mills theory and holographic models of QCD, and discuss its range of applicability.

\end{abstract}


\maketitle

\section{Introduction and summary}

One important application of the AdS/CFT correspondence \cite{Maldacena:1997re, Aharony:1999ti} is the calculation of transport properties such as viscosity and  conductivity of strongly coupled large $N_c$ gauge theory plasmas using methods from black hole physics. This follows from the map of thermal states of field theories to classical gravity solutions containing black holes. Thus, certain intractable strongly coupled field theory calculations are reduced to the study of black hole perturbations in General Relativity.

One such example is the gravitational calculation of the ratio of the shear viscosity $\eta$ to the entropy density $s$ which yields the value
$\frac{1}{4\pi}$  \cite{Policastro:2001yc}, and is generic for strongly coupled gauge theories with an Einstein gravity holographic dual (see e.g. \cite{Buchel:2003tz}). On the experimental side, it has been argued that the value of $\eta/s$ in  the quark-gluon plasma produced in Relativistic Heavy Ion Collisions is indeed of the same order of magnitude (see e.g. \cite{Romatschke:2007mq}).

The $\frac{1}{4\pi}$ result can be obtained using a number of inter-related calculations. In \cite{Policastro:2001yc} a holographic calculation
of the two-point correlator of the stress-energy tensor was performed and the shear viscosity was extracted from it using the Kubo formula.
The transport coefficients can be extracted also from the linearized quasi-normal modes on the black hole background, which in the hydrodynamic limit, correspond to shear and sound modes of the gauge theory \cite{qnms}.

In \cite{membranefluct} it was shown that the diffusive modes could be understood in terms of fluctuations of the black hole horizon using the language of the membrane paradigm \cite{membrane}, where one identifies a horizon with a fictitious viscous fluid. In \cite{Iqbal:2008by} this approach to calculations of shear viscosity and conductivity was formalized, showing how and when the horizon captures the response of the boundary theory.

The hydrodynamics of non-conformal gauge theories is also characterized by a bulk viscosity coefficient $\zeta$. In general, holographic bulk viscosities do not exhibit the simple universal character of the shear viscosity. Also, the status of bulk viscosity in the membrane paradigm is more mysterious. In the old membrane paradigm  \cite{membrane}, bulk viscosity is identified as negative, with a ratio $\zeta/\eta = -1$. Since the bulk viscosity of a real non-conformal fluid must be positive (and generally also non-universal), it has been expected that a horizon based calculation will never correctly capture the bulk viscosity. The analysis of shear viscosity  in \cite{Iqbal:2008by} relies on the simplicity of the equation satisfied by the relevant transverse graviton mode $h_x^y$ on the black hole background.  To compute the bulk viscosity, one must instead consider the trace mode $h_i^i$.  In the gravity solutions dual to non-conformal gauge theories, this mode may be mixed with fluctuations of scalar fields, and the calculations will be much more involved.

The purpose of this paper is to study how the horizon captures the bulk viscosity in the framework of the fluid/gravity correspondence.   In the conformal setup one considers a boosted black brane solution, characterized by uniform boost velocity $u^\mu$, Hawking temperature $T$, and also, in general, chemical potentials $\mu_a$ and the corresponding charge densities $\rho^a$. This is dual to an equilibrium perfect fluid state of the boundary gauge theory. Next, one perturbs this solution by making the ansatz that the fluid velocity and thermodynamic variables are spacetime functions and solving the Einstein equations in a derivative expansion in Knudsen number $\ell_{mfp}/L \ll 1$, where $\ell_{mfp}$ is the mean free path and $L$ the size of perturbations \cite{Bhattacharyya:2008jc}. With the solution to first order in derivatives, one can compute the holographic boundary stress tensor at the first viscous order and read off the viscosities. The constraint equations projected on the boundary are the relativistic Navier-Stokes equations.

In \cite{GaussCodazzi} it was shown that in these solutions the constraint Einstein equations projected on the horizon (Gauss-Codazzi equations) capture the same information in a simple way. In particular, the null focusing (Raychaudhuri) equation describing the evolution of the horizon entropy is equivalent to the viscous fluid entropy balance law. Using the zeroth order solution in derivatives one can read off the famous $\frac{1}{4\pi}$ ratio and thermal conductivities of holographic abelian and non-abelian gauge theories as well \cite{charges} .
The various schemes are related by an RG flow in the radial direction
\cite{Bredberg:2010ky}.

In this paper we will extend this method to non-conformal fluids. We will show that the bulk viscosity arises from the flux of scalar fields across the null horizon and derive a novel and simple formula for the bulk viscosity in terms of derivatives of the horizon value(s) of the scalar field(s) with respect to thermodynamic variables such as the entropy and charge densities.

We will consider  $(d+1)$-dimensional gravitational backgrounds holographically describing thermal states in  strongly coupled $d$-dimensional field theories. The
$(d+1)$-dimensional gravitational action in the Einstein frame reads
\begin{align}
I = \frac{1}{16\pi} \int \sqrt{-g} d^{d+1} x \left({\cal R} - \frac{1}{2} \sum_i (\partial \phi_i)^2 - V(\phi_i)\right) + I_{gauge}  \ ,
\label{ac}
\end{align}
where we set $\hbar = c = G^{d+1}_N=1$ and canonically normalized the kinetic terms for the various scalar fields $\phi_i$. $V(\phi_i)$ represents the potential for the scalar fields and $I_{gauge}$ represents the action of gauge fields (abelian or non-abelian) $A_{\mu}^a$.

The null horizon focusing equation obtained by projecting the field equations of (\ref{ac}) on the horizon is equivalent via the fluid/gravity correspondence to the entropy balance law of the fluid. Using this equation we will derive the following formula for the fluid bulk viscosity $\zeta$:
\begin{align}
\frac{\zeta}{\eta} = \sum_i \left(s \frac{d \phi^{H}_i}{ds} + \rho^a \frac{d \phi^{H}_i}{d\rho^a} \right)^2 \ . \label{ratio}
\end{align}
Here $\eta$ is the shear viscosity, $s$ is the entropy density, $\rho^a$ are the charges associated with the gauge fields $A_{\mu}^a$, and $\phi_i^H$ are the values of the scalar fields on the horizon. We will apply (\ref{ratio}) to three general classes of gauge theory plasmas: non-conformal branes, perturbations of ${\cal N}=4$ supersymmetric Yang-Mills (SYM) and holographic models of QCD.

The organization of this paper is as follows. In Section II  we will briefly describe the horizon geometry and how the (Raychaudhuri) null focusing equation is mapped to the equation of the fluid entropy law. In a generic setup, we will show how the bulk viscosity at the horizon is encoded in the simple formula. In Section III we will apply the formula to the hydrodynamics of non-conformal theories such as those described by the near-horizon limit of Dp-branes, the Sakai-Sugimoto model in the quenched
approximation, and non-conformal models that arise via a compatification of a higher dimensional conformal theory.  The formula correctly reproduces the results of  \cite{Mas:2007ng}, \cite{Benincasa:2006ei} and  \cite{Kanitscheider:2009as}. We then consider in Section IV deformations of $\mathcal{N} = 4$ SYM. These include the $\mathcal{N} = 2^{\star}$ plasma studied in \cite{Benincasa:2005iv}, \cite{Buchel:2008uu}, a deformation of charged $\mathcal{N} = 4$ plasma studied in
\cite{Buchel:2010gd} and the  softly broken conformal symmetry theories studied in  \cite{Yarom:2009mw}. Our formula will reproduce the results of these papers. In particular, we can obtain analytical results for the bulk to shear viscosity ratio without the need to resort to numerical methods.  In Section V we will consider the bulk viscosity of non-critical backgrounds proposed as phenomenological holographic descriptions of QCD \cite{Gubser:2008sz,Kiritsis}. We will show that our formula captures the bulk to shear viscosity ratio of these models in either the high temperature limit or in the adiabatic approximation, but otherwise deviates. In the last section we conclude with a discussion of the formula's range of applicability.

\section{Bulk viscosity from the null focusing equation }

In this section we will start with a brief review of the horizon geometry and of the derivation of the null focusing (Raychaudhuri) equation \cite{GaussCodazzi}. Using the fluid/gravity correspondence, we map this equation into the fluid entropy law. We then show how the bulk viscosity is encoded in a novel and simple formula.

\subsection{Horizon geometry and the focusing equation}
\label{geometry}

We denote the coordinates of the bulk spacetime  by $x^A = (r, x^\mu), A=0,...,d$. The $x^\mu$ are local coordinates on the horizon $\mathcal{H}$; $r$ is a transverse coordinate, with $r = 0$ on $\mathcal{H}$. $\del_A r$ is a null covector tangent to the $\mathcal{H}$. When raised with the bulk metric, it gives a vector field $\ell^A$ %
\begin{align}
\ell^A = g^{AB}\del_B r = 	(0, \ell^\mu) \ .
\end{align}
These choices fix the following components of the inverse bulk metric on $\mathcal{H}$:
\begin{align}
	g^{rr} = 0; \quad g^{r\mu} = \ell^\mu \ . \label{eq:g^rA}
\end{align}

The pullback of $g_{AB}$ into $\mathcal{H}$ is the degenerate horizon metric $\gamma_{\mu\nu}$. Its null directions are the generating light-rays of $\mathcal{H}$, i.e. $\gamma_{\mu\nu}\ell^\nu = 0$. The Lie derivative of $\gamma_{\mu\nu}$ along $\ell^\mu$ gives us the shear/expansion tensor, or ``second fundamental form'':
\begin{align}
	\theta_{\mu\nu} \equiv \frac{1}{2}\mathcal{L}_{\ell}\gamma_{\mu\nu} \ . \label{eq:theta_def}
\end{align}
We can write a decomposition of $\theta_{\mu\nu}$ into a shear tensor $\sigma^{(H)}_{\mu\nu}$ and an expansion coefficient $\theta$:
\begin{align}
	\theta_{\mu\nu} = \sigma^{(H)}_{\mu\nu} + \frac{1}{d-1}\theta\gamma_{\mu\nu} \ . \label{eq:theta_decompose}
\end{align}
The expansion $\theta$ can be expressed as
\begin{align}
\theta = v^{-1} \partial_\mu (v \ell^\mu) = v^{-1} \partial_\mu S^{\mu}\ ,
\end{align}
where $v$ is a scalar density equal to the horizon area density,  and $S^\mu = v l^{\mu}$ is the horizon area current \cite{Bhattacharyya:2008xc}.

It is convenient to raise indices with $(G^{-1})^{\mu\nu}$, which is the inverse of any matrix $G_{\mu\nu}$ of the form \cite{charges}
\begin{align}
	G_{\mu\nu} = \lambda\gamma_{\mu\nu} - b_\mu b_\nu; \quad b_\mu\ell^\mu \neq 0 \ . \label{eq:G}
\end{align}
Here we introduced the scalar field $\lambda$ for later convenience: it will turn out that a matrix of the form \eqref{eq:G} coincides at leading order with the metric $h_{\mu \nu} = \eta_{\mu\nu}$ associated with the (typically flat) metric of the hydrodynamic dual theory.

Since $\gamma_{\mu\nu}$ is degenerate, one cannot use it to define an intrinsic connection on the null horizon, as could be done for spacelike or timelike hypersurfaces. The bulk spacetime's connection does induce a notion of parallel transport in $\mathcal{H}$, but only along its null generators. This structure is not fully captured by $\gamma_{\mu\nu}$; instead, it is encoded by the extrinsic curvature, or 'Weingarten map' $\Theta_\mu{}^\nu$, which is the horizon restriction of $\nabla_A\ell^B$:
\begin{align}
	\Theta_\mu{}^\nu = \nabla_\mu\ell^\nu \ .
\end{align}
One can show that given an arbitrary $G_{\mu\nu}$ of the form \eqref{eq:G}, $\Theta_\mu{}^\nu$ can be written as \cite{charges}:
\begin{align}
	\Theta_\mu{}^\nu = \lambda\theta_{\mu\rho}(G^{-1})^{\rho\nu} + c_\mu\ell^\nu; \quad c_\mu\ell^\mu = \kappa \ . \label{eq:Theta}
\end{align}
$\kappa$ is the ``surface gravity",  which measures the non-affinity of $\ell^\mu$. The covector $c_\mu$ encodes the degrees of freedom in $\Theta_\mu{}^\nu$ which are independent of $\gamma_{\mu\nu}$.  In the hydrodynamics, these degrees of freedom will roughly correspond to the velocity and temperature fields.

Re-expressing the identity,
\begin{align}
{\cal R}_{AB} \ell^A \ell^B = \ell^A (\nabla_B \nabla_A - \nabla_A \nabla_B ) \ell^B \ ,
\end{align}
in terms of $\Theta_\mu{}^\nu$ and imposing the Einstein field equation, we get the null focusing (Raychaudhuri) equation
\begin{align}
	 -\ell^\mu\del_\mu\theta + \kappa\theta - \frac{1}{d-1}\theta^2
		- \lambda^2(G^{-1})^{\mu\rho}(G^{-1})^{\nu\sigma}\sigma^{(H)}_{\mu\nu}\sigma^{(H)}_{\rho\sigma} = 8\pi T_{AB} \ell^A \ell^B \ , \label{eq:focusing_basic}
\end{align}
where $T_{AB}$ is the bulk matter stress tensor.

\subsection{Fluid entropy law and bulk viscosity}
\label{focusingfluid}

In the following, we will consider gravitational solutions holographically dual to  thermal states in strongly coupled field theories.
We will begin with the action (\ref{ac}); its field equations read
\begin{align}
{\cal R}_{AB} - \frac{1}{2} g_{AB} {\cal R} = \frac{1}{2} \sum_i \left(\partial_A \phi_i \partial_B \phi_i + \frac{1}{2} g_{AB} V(\phi_i)\right)+ T^{gauge}_{AB} \ .
\end{align}
In (radially shifted) Eddington-Finkelstein coordinates the metric solutions that we are interested in  take the form
\begin{align}
g_{AB} dx^A dx^B  = ds^2_{(0)} =  -c^2_T(r+R) \ell_\mu \ell_\nu dx^\mu dx^\nu  + 2 c_R(r+R) \ell_\mu dx^\mu dr + c^2_X(r+R) P_{\mu \nu} dx^\mu dx^\nu, \label{zerothmetric}
\end{align}
where at $r=0$, the function $c_T$ vanishes and there is a horizon. The vector $\ell^\mu$ is a boost vector and has the dual role as a fluid velocity and the null normal to the horizon at $r=0$. The indices $\mu$, $\nu$ etc. are raised with the flat metric $G^{(0)}_{\mu \nu} = \eta_{\mu \nu}$, and
\begin{align}
P_{\mu \nu} = \eta_{\mu \nu} + \ell_\mu \ell_\nu \ ,
\end{align}
 is the projection tensor. In addition there are the solutions to the scalar fields  and gauge fields
\begin{align}
\phi_i(r+R),~~~~~~~~A^a_B = A^a(r) \delta^t_B \ .
\end{align}
The Bekenstein-Hawking entropy density and the Hawking temperature associated with the metric are
\begin{align}
s= \frac{1}{4} c_X(R)^{d-1}; \quad T = \frac{c_T(R) c'_T(R)}{c_R(R)} \ .
\end{align}
From the solution for the gauge field, one can read off the values of the chemical potentials $\mu_a$ and the charge densities $\rho^a$.
The energy density and pressure associated with this solution are determined from the formulas
\begin{align}
d\epsilon = T ds + \mu_a d\rho^a,~~~~~~\epsilon+P = sT+\mu_a \rho^a \ .
\end{align}

With the thermodynamic quantities in hand, we can follow the ansatz of the fluid-gravity correspondence, where one allows the thermodynamic variables, $R$, and $\ell^\mu$ to be functions of $x^\mu$. Now the zeroth order metric \eqref{zerothmetric}  and other fields as functions of $x^\mu$ are no longer solutions and must be corrected order by order in a derivative expansion. Let us see what we can learn from the Raychaudhuri equation (\ref{eq:focusing_basic}). We write it in the following form corresponding to the action \eqref{ac}
\begin{align}
 -\ell^\mu\del_\mu\theta + \kappa\theta - \frac{1}{d-1}\theta^2- \lambda^2(G^{-1})^{\mu\rho}(G^{-1})^{\nu\sigma}\sigma^{(H)}_{\mu\nu}\sigma^{(H)}_{\rho\sigma} = \frac{1}{2} \sum_i (D \phi_i)^2 + 8\pi T^{gauge}_{\mu \nu} \ell^\mu \ell^\nu, \label{focusing}
\end{align}
where we define the operator $D \equiv \ell^\mu \partial_\mu$. Note, that $V(\phi_i)$ does not appear in the equation since $\ell^\mu$ is null.
In the equilibrium solution, the expansion and shear vanish identically, while the scalars and gauge fields parts vanish since these fields have only a radial dependence.  Consider now the focusing equation at first order in derivatives of $x^\mu$. In the case where $T^{gauge}_{\mu \nu}$ are general non-abelian gauge fields, one can show \cite{GaussCodazzi,charges} that the first contribution appears at $O(x^2)$. Hence the focusing equation at order $O(x)$ reduces to
\begin{align}
\theta^{(1)} = \partial_\mu (s \ell^\mu) =0 \ , \label{idealentropy}
\end{align}
which matches the ideal entropy conservation law for the fluid. Here the index on $\theta$ refers to the order of the term in the derivative expansion.

In proceeding to second order $O(x^2)$, one may worry that we might need information about the first order corrected metric. However, the structure of the Raychaudhuri equation is such that this is not the case.  Consider first the shear term. The first order horizon shear can be calculated from the zeroth order metric. With the choice $\lambda = (4s)^{\frac{-2}{d-1}}$ the shear squared term becomes
\begin{align}
\lambda^2(G^{-1})^{\mu\rho}(G^{-1})^{\nu\sigma}\sigma^{(H)}_{\mu\nu}\sigma^{(H)}_{\rho\sigma} = \pi^{\mu \nu} \pi_{\mu \nu} \ ,
\end{align}
where
\begin{align}
\pi_{\mu \nu} = P^\rho_\mu P^\sigma_\nu \partial_{(\rho}\ell_{\sigma)} - \frac{1}{d-1}P_{\mu\nu} \partial_\rho \ell^\rho \ .
\end{align}

Next, the first three terms of \eqref{focusing} to this order are:
\begin{align}
\ell^\mu \partial_\mu \theta^{(1)} + \kappa^{(0)} \theta^{(1)} + \kappa^{(1)} \theta^{(1)} + \kappa^{(0)} \theta^{(2)} - \frac{1}{d-1}\theta^2_{(1)} \ .
\end{align}
However, at this order  we can substitute the ideal equation $\theta^{(1)}=0$. Also, at first order there is always an ambiguity in the definition of our variables $\ell^\mu$ and $s$ (or $T$). We remove this by the frame choice that $\ell^\mu$ always coincides with the entropy current, that is that $S^{(1) \mu} = 0$ and $s^{(1)} =0$. Therefore the $\theta^{(2)}$ piece also vanishes.

In \cite{charges} it was shown that any gauge field contribution at $O(x^2)$ can be expressed in terms of (positive definite) entropy production terms associated with conductivities. From here on, we neglect this term and focus on the viscosities. Using $\kappa^{(0)} =2\pi T$, the focusing equation \eqref{focusing} has the form
\begin{align}
\partial_\mu (s \ell^\mu) = \frac{1}{4} \partial_\mu S^\mu = \frac{s}{2\pi T} \pi_{\mu \nu} \pi^{\mu \nu} + \frac{s}{4\pi T}\sum_i (D \phi_i)^2 + \cdots \label{focusing1}
\end{align}
where the dots indicate the gauge field terms.
Notice that the scalar term is a derivative of $\phi^{(0)}_i(r,x^\mu)$. We take the derivative of this solution with respect to $x^\mu$ and then evaluate at $r=0$. This is equivalent to $D \phi^{H}_i(x^\mu)$ of the horizon values. The horizon values $\phi^{H}_i$ are functions of the entropy density $s(x^\mu)$ and charge density $\rho^a(x^\mu)$ (or the Hawking temperature $T(x^\mu)$ and chemical potential $\mu_a(x^\mu)$). Thus we can write
\begin{align}
D \phi^{H} = \frac{d \phi^{H}_i}{ds} Ds +  \frac{d \phi^{H}_i}{d\rho^a} D\rho^a = - \left(s \frac{d \phi^{H}_i}{ds} + \rho^a \frac{d \phi^{H}_i}{d\rho^a} \right) (\partial_\rho \ell^\rho) \label{derivativeformula} \ .
\end{align}
In the last equality we used the ideal conservation law \eqref{idealentropy} and also the ideal conservation equation for any charge current present
\begin{align}
\partial_\mu J^{\mu a} = \partial_\mu (\rho^{a} \ell^\mu) =0 \ .
\end{align}
Therefore, the Raychaudhuri equation to second order takes in the fluid/gravity correspondence the form of a fluid entropy balance law \cite{Landau},
\begin{align}
\partial_\mu (s \ell^\mu) = \frac{2\eta}{T} \pi_{\mu \nu} \pi^{\mu \nu} + \frac{\zeta}{T} (\partial_\rho \ell^\rho)^2 \ ,
\end{align}
with the standard shear viscosity obeying $\eta/s = 1/4\pi$, but now with an additional bulk viscosity term.
Substituting \eqref{derivativeformula} into \eqref{focusing1}  we find the bulk to shear viscosity ratio implied by the Raychaudhuri equation is simply
\begin{align}
\frac{\zeta}{\eta} = \sum_i \left(s \frac{d \phi^{H}_i}{ds} + \rho^a \frac{d \phi^{H}_i}{d\rho^a} \right)^2 \ ,
\end{align}
or equivalently
\begin{align}
\frac{\zeta}{s} = 4 \pi \sum_i \left(s \frac{d \phi^{H}_i}{ds} + \rho^a \frac{d \phi^{H}_i}{d\rho^a} \right)^2 \ .
\end{align}

In the absence of charges we can use  the speed of sound $v_s^2 = \frac{d(\ln T)}{d(\ln s)}$, and recast the formula in the form
\begin{align}
\frac{\zeta}{\eta} = \sum_i v_s^4 T^2 \left(\frac{d\phi_i}{dT}\right)^2 \label{temp} \ .
\end{align}

\section{Non-conformal branes}

In this section we will use the formula (\ref{ratio}) to calculate the bulk viscosity of non-conformal hydrodynamic models realized holographically as gravitational backgrounds with scalars, dual to non-conformal branes. We will consider three examples: Dp-brane here $p\neq 3$ \cite{Mas:2007ng,Benincasa:2006ei}, the Sakai-Sugimoto model in the quenched approximation \cite{Benincasa:2006ei} and non-conformal brane backgrounds obtained from the dimensional reduction of pure gravity \cite{Kanitscheider:2009as}.

We will show that in all the three models, the horizon formula (\ref{ratio}) reproduces correctly the previous results obtained
by either the holographic hydrodynamics or by a holographic calculation of the stress-energy two-point correlators and the use of
the Kubo formula (or the equivalent quasi-normal modes analysis).

\subsection{Dp-branes}

 The decoupling limit of stacks of coincident Dp branes  with $p \neq 3$ yields a gauge/gravity correspondence of non-conformal gauge theories \cite{Itzhaki:1998dd}. The (super)gravity solution is conformal to $AdS_{p+2} \times S^{8-p}$, and
a key feature of the solution is the existence of a non-trivial scalar dilaton field.

The $(p+2)$-dimensional action for the metric and dilaton, reduced from the ten-dimensional supergravity action reads
\begin{align}
S =  \frac{1}{16\pi}\int d^{p+2} x \sqrt{-g} \left({\cal{R}} - \frac{1}{2} (\partial\phi)^2 - \frac{(7-p)(9-p)}{2} e^{\frac{2(3-p)}{\sqrt{2 p (9-p)}}\phi(r)}\right) \ ,
\end{align}
where we canonically normalized the scalar kinetic term.
The zeroth order equilibrium solution for the metric components can be put in the form of the metric \eqref{zerothmetric} with
\begin{align}
c^2_T(r) =r^{\frac{9-p}{p}}f(r);  \quad c^2_X(r) = \left(r+R\right)^{\frac{9-p}{p}}; \quad c_R(r) = r^{\frac{(p-3)(p-6)}{2p}} , \label{metricsolution}
\end{align}
where $f(r) = (1-\frac{R}{r+R})^{7-p}$.  For the scalar dilaton, the solution is
\begin{align}
\phi(r) =\frac{\sqrt{2}}{2} (3-p) \left(\frac{9-p}{p}\right)^{\frac{1}{2}} \ln(r+R) \label{dilaton} \ .
\end{align}

The thermodynamics of this equilibrium solution is characterized by the Hawking temperature and the Bekenstein-Hawking area entropy density
\begin{align}
T = \frac{\kappa}{2\pi} = \frac{7-p}{4\pi} R^{(5-p)/2}; \quad s = \frac{v}{4} = (1/4) R^{(9-p)/2} \ .
\end{align}
The value of the dilaton on the horizon as a function of the entropy density is
\begin{align}
\phi^H = \frac{\sqrt{2}}{2} (3-p) \left(\frac{9-p}{p}\right)^{\frac{1}{2}} \ln((4s)^{\frac{2}{9-p}}) \ . \label{equilib}
\end{align}
Taking the derivative of $\phi^H$ with respect to the entropy density, we find
using \eqref{ratio}
\begin{align}
\frac{\xi}{\eta} = \frac{2(3-p)^2}{p (9-p)},
\end{align}
which agrees with  \cite{Mas:2007ng}, who computed the ratio using the linearized quasi-normal modes analysis of this background solution.

\subsection{The Sakai-Sugimoto model}

We now consider the hydrodynamics of the Sakai-Sugimoto model \cite{Sakai:2004cn} proposed as a holographic model of QCD with fundamental flavours. We will consider the quenched approximation discussed in \cite{Benincasa:2006ei}. The effective five-dimensional action reads
\begin{align}
S = \frac{1}{16\pi} \int \sqrt{-g} d^5 x \left({\cal{R}} - \frac{1}{2} (\del f)^2 - \frac{1}{2} (\del \Phi)^2- \frac{1}{2} (\del w)^2-\mathcal{P}\right) \ ,
\end{align}
where $\mathcal{P}$ is a potential function that depends on the three scalars $f,\Phi,w$. The zeroth order solution again can be put in the same form as \eqref{zerothmetric}. The entropy density is
\begin{align}
s =\frac{1}{4} R^{\frac{5}{2}} \ ,
\end{align}
and the zeroth order scalar field profiles are
\begin{align}
\Phi = \frac{3}{4} \ln (r+R), \quad w =  \frac{\sqrt{10}}{5} \ln (r+R), \quad f =  \frac{13}{4\sqrt{15}} \ln (r+R) \ .
\end{align}
Evaluating at the horizon and using the formula \eqref{ratio} we find
\begin{align}
\frac{\xi}{\eta} = \frac{4}{25} \left(\frac{9}{16}+\frac{10}{25}+\frac{169}{240}\right) = \frac{4}{15}  \ ,
\end{align}
which agrees with the result found in \cite{Benincasa:2006ei}.

\subsection{Non-conformal branes}

It has been demonstrated in  \cite{Kanitscheider:2009as}  that solutions which asymptote locally to non-conformal branes can be obtained from higher-dimensional asymptotically AdS solutions via a dimensional reduction and a continuation in dimension.  This generates a class of holographic descriptions
of $d$-dimensional non-conformal
hydrodynamic models parametrized by positive number $\sigma$, where the energy density $\varepsilon$ and the pressure $p$ are related by the equation of state $\varepsilon = (2\sigma -1) p$, where
$2\sigma \neq d$ (recall that $\varepsilon = (d-1)p$ is the equation of state for conformal hydrodynamics).
By using the hydrodynamics expansion, one can find a general formula for the ratio of the bulk to the shear viscosity in this class of models, it reads \cite{Kanitscheider:2009as}
\begin{align}
\frac{\zeta}{\eta} = \frac{2(2\sigma-d)}{(d-1)(2\sigma-1)} \ . \label{NC}
\end{align}
We will now derive this result using (\ref{ratio}).

In \cite{Kanitscheider:2009as} the authors work in the dual frame, where the action is
\begin{align}
 S = \frac{1}{16\pi}\int d^{d+1} x \sqrt{-g} e^{\phi} \left({\cal{R}} + \frac{2\sigma-d-1}{2\sigma-d} (\partial \phi)^2 + 2\sigma(2\sigma-1)\right) \label{ksaction} \ .
\end{align}
The corresponding black brane solution reads
\bea
ds^2 &=& (r+R)^2 P_{\mu \nu} dx^\mu dx^\nu - f(r) \ell_\mu \ell_\nu dx^\mu dx^\nu + 2 \ell_\mu dx^\mu dr \\
\phi &=& (2\sigma-d) \ln (r+R) \ ,
\eea
where $f(r)= 1-\left(\frac{r+R}{R}\right)^{-2\sigma}$ and $R= \frac{2\pi T}{\sigma}$.

The entropy density in the dual frame is \emph{not} the area density $v$, instead we have
\begin{align}
s = \frac{1}{4} \exp(\phi^H) v \ ,
\end{align}
which for this solution is
\begin{align}
s = \frac{1}{4} R^{2\sigma-1} \label{dualentropy} \ .
\end{align}
Our analysis of the Raychaudhuri equation in Section II B depends on the entropy density being proportional to an area density. Therefore, to apply our formula \eqref{ratio}, we need to make a conformal transformation of \eqref{ksaction} to the Einstein frame. The bulk viscosity of course does not depend on the choice of conformal frame. In the Appendix we re-analyze the Raychaudhuri equation in the dual frame and arrive at the same result.

To transform the action \eqref{ksaction} to the form of \eqref{ac}, we first make the conformal transformation
\begin{align}
\tilde{g}_{AB} = e^{\frac{2\phi}{d-1}} g_{AB},
\end{align}
where the tilde represents the metric in the Einstein frame, and then canonically normalize the scalar field. On the horizon, the scalar field solution then has the form
\begin{align}
\phi^{H} = \frac{\sqrt{2}\sqrt{(2\sigma-1)(2\sigma-d)}}{\sqrt{d-1}} \ln (R).
\end{align}
Re-expressing this in terms of the entropy density, \eqref{dualentropy}, and using the formula (\ref{ratio}) we obtain (\ref{NC})
in agreement with the result found in \cite{Kanitscheider:2009as}.

\section{Perturbations of $\mathcal{N}=4$ plasma}

In this section we consider deformations of $\mathcal{N} = 4$ SYM. These include the $\mathcal{N} = 2^{\star}$ plasma studied in \cite{Benincasa:2005iv}, \cite{Buchel:2008uu}, a deformation of charged $\mathcal{N} = 4$ plasma studied in
\cite{Buchel:2010gd} and the  softly broken conformal symmetry theories studied in  \cite{Yarom:2009mw}. We will show that
the bulk viscosity analytical formula agrees with the previous calculations, some of which required previously a numerical analysis.

\subsection{$\mathcal{N}=2^{\star}$ plasma}

The $\mathcal{N}=2^{\star}$ four-dimensional gauge theory is defined as a mass deformation $\mathcal{N} = 4$ SYM, where
the bosons and fermions masses $m_b$ and $m_f$ break the conformal symmetry.
The $\mathcal{N}=2^{\star}$ gauge theory has, in the limit of large t'Hooft coupling, an explicit supergravity dual description called the Pilch-Warner flow \cite{Pilch:2000ue}.  The five-dimensional geometry dual to a finite temperature state is characterized by three independent parameters: the temperature and the horizon values of two scalar fields $\alpha$ and $\chi$.

The effective five-dimensional action for this model is given by \cite{Benincasa:2005iv}
\begin{align}
S = \frac{1}{16\pi} \int \sqrt{-g} d^5 x \left( {\cal {R}} - \frac{1}{2} (\del \alpha)^2 - \frac{1}{2} (\del \chi)^2- 4\mathcal{P}\right) \ ,
\end{align}
where we canonically normalized the scalar fields $\alpha$ and $\chi$ and $\mathcal{P}$ is their potential. The values of
the scalar fields at the horizon are denoted by \cite{Benincasa:2005iv}
\begin{align}
\alpha = \delta_1; \quad \chi= \delta_2; \quad c_X = \delta_3 \ .
\end{align}
In the high temperature regime of the $\mathcal{N}=2^{\star}$ black brane, the gravity field equations can be solved analytically in an expansion in parameters $\delta_i \ll 1$ \cite{Buchel:2003ah, Benincasa:2005iv} with
\begin{align}
\delta_1 = -\frac{1}{\sqrt{24} \pi} \left(\frac{m_b}{T}\right)^2, \quad \delta_2 = \frac{\sqrt{2} \Gamma(3/4)^2}{\pi^{3/2}} \frac{m_f}{T}, \quad 2\pi T = \delta_3(1+\frac{2}{3\pi^2}\delta_1^2+\frac{1}{6\pi}\delta_2^2)  \ . \label{horizondata}
\end{align}
The entropy density associated with this solution is $s = \delta_3^3/4$.
Plugging the derivatives of the horizon values of the scalars with respect to the temperature in (\ref{temp})
we get
\begin{align}
\frac{\zeta}{\eta} = v_s^4 (72 x + 6 y) \ ,
\end{align}
where we defined
\begin{align}
y = \frac{\Gamma(3/4)^4}{3\pi^3} \frac{m_f^2}{T^2}; \quad x= \frac{1}{432\pi^2} \frac{m_b^4}{T^4} \ .
\end{align}
Since we work at linear order, we can use the zeroth order (conformal) value for the speed of sound, $v_s = \frac{1}{\sqrt{3}}$,
and arrive at
\begin{align}
\frac{\zeta}{\eta} = 8 x + \frac{2}{3} y \ .
\end{align}
Comparing to the parametrization  $\zeta/\eta =\beta^{\Gamma}_b  x + \beta^{\Gamma}_f y$ in \cite{Benincasa:2005iv}, we see that
$\beta^{\Gamma}_b = 8$ and $\beta^{\Gamma}_f =2/3$.  This is consistent with the values reported in \cite{Buchel:2008uu}, where $\beta^{\Gamma}_b \sim 8.001$ and $\beta^{\Gamma}_f \sim 0.6666$ were obtained via numerical analysis of quasi-normal sound modes.

\subsection{Deformation of charged $\mathcal{N}=4$ plasma}

The previous examples that we considered did not involve charges.
An interesting test case for our formula is the mass deformation of the charged $\mathcal{N}=4$ plasma studied recently in \cite{Buchel:2010gd}.  The action for this theory is given by
\begin{align}
S = \frac{1}{16\pi} \int d^5 x \sqrt{-g} \left({\cal R} - \frac{1}{4} \phi^{4/3} F^2 - \frac{1}{3} \phi^{-2} (\partial \phi)^2 + 4 \phi^{2/3} + 8 \phi^{-1/3} + \delta L\right) \ ,
\end{align}
where $\delta L$ represents the conformal symmetry breaking mass deformation due to a second scalar field $\chi$,
\begin{align}
\delta L = -\frac{1}{2} (\partial \chi)^2 - \frac{m^2}{2} \chi^2 + \cdots
\end{align}

The idea of the calculation is to work perturbatively in the $\chi$ field. One makes the ansatz that $\chi = \lambda \chi_1$, where $\lambda$ is the coefficient of the non-normalizable mode of $\chi$. The gauge field $A_B$ and metric functions receive corrections at $O(\lambda^2)$. At zeroth order in $\lambda$ (the unperturbed solution), the thermodynamic variables are defined in terms of constants $\kappa$ and $\beta$:
\begin{align}
2\pi T = \beta \frac{\kappa+2}{\sqrt{\kappa(1+\kappa)}}; \quad \mu = \frac{\beta}{\sqrt{1+\kappa}}; \quad \frac{2\pi T}{\mu} = \sqrt{\kappa}+\frac{2}{\sqrt{\kappa}} \ . \label{horizondata2}
\end{align}
The ratio of the temperature to the chemical potential obtains a minimum at $\kappa=2$, where the black hole undergoes a second order phase transition. The scalar field $\phi$ has the horizon value
\begin{align}
\phi^H = 1 + \kappa \ .
\end{align}

The zeroth order solution should have a vanishing bulk viscosity, but the scalar field $\phi$ has a non-trivial horizon value, which seems to imply a potential contradiction. Let's consider the general formula \eqref{ratio}. The Eq. (2.47) in \cite{Buchel:2010gd} for the entropy and charge density imply that
\begin{align}
\frac{\rho^2}{s^2} = \frac{\kappa}{4\pi^2} \ . \label{rhos}
\end{align}
Using (\ref{rhos}) we can express $\phi^H$ as a function of $\rho$ and $s$, and it is straightforward to see that the formula
for the bulk viscosity (\ref{ratio}) yields $\zeta=0$ as expected from conformal invariance at this order.

Non-zero bulk viscosity arises at the order $O(\lambda^2)$ from the horizon value of the $\chi$ field, where
the parameter $\lambda$ is related to the fermionic mass $M$ and the temperature by
\begin{align}
\lambda = \frac{(\kappa+2)}{2^{3/4} (1+\kappa)^{3/4} \pi} \frac{M}{T} \ .
\end{align}
Thus the expansion in $\lambda$ is again an expansion in the limit of high temperature, where $T \gg M$.
The field $\phi$ does not contribute at this order since its form changes at order $O(\lambda^2)$ and therefore its contribution
to the bulk viscosity is at order $O(\lambda^4)$.

We can compare to the numerical calculation done in \cite{Buchel:2010gd} at the critical point ($\kappa=2$). The field $\chi$
takes the form
\begin{align}
\chi^H = \lambda c^{h (0)}_0 \ .
\end{align}
The coefficient $c^{h (0)}_0 = 0.7464562054847809$ (see Table I in \cite{Buchel:2010gd})  is a pure number obtained numerically by solving ordinary differential equations subject to boundary conditions at the horizon and at infinity.

When using the formula (\ref{ratio}) we hold $\kappa$ fixed and vary the temperature $T$ with respect to the entropy density $s$ and the charge density $\rho$.
Using the relation between $T, s, \rho$ in Eq. (2.47) in \cite{Buchel:2010gd} is easy to see
that (\ref{ratio}) gives
\begin{align}
\frac{\zeta}{\eta} = \frac{1}{9} (c^{h (0)}_0)^2 \lambda^2 \sim 0.0619 \lambda^2 \ ,
\end{align}
which is consistent with the numerical result of $\sim 0.06218 \lambda^2$ obtained by analysis of the quasi-normal sound modes \cite{Buchel:2010gd}.

\subsection{Softly broken conformal symmetry}

A general formula for bulk viscosity of  theories, where the conformal symmetry is softly broken has been derived in \cite{Yarom:2009mw}.  One considers a bulk gravitational action with a scalar field $\phi$,
\begin{align}
S= \frac{1}{16\pi} \int \sqrt{-g} \left({\cal R} - (d-1)d - \frac{1}{2} (\partial \phi)^2 + V(\phi) \right) d^{d+1} x \ .
\end{align}
The conformal symmetry of the field theory is broken by a relevant operator $\mathcal{O}_{\Delta}$ of conformal dimension $\Delta$. In the duality prescription, the field
$\phi$ is dual to $\mathcal{O}_{\Delta}$; the expectation value of $\mathcal{O}_{\Delta}$ and the source term $\Lambda^{d-\Delta}$ can be read off from a boundary series expansion of $\phi$.

In the high temperature $\Lambda/T \ll 1$ expansion, the relevant physics is simply captured by the scalar field propagating on the fixed black brane background.
 For our purposes, all we need is the (zeroth order) scalar field solution
\begin{align}
\phi(rb) = (\Lambda b)^{d-\Delta} \frac{2 \Gamma(\frac{\Delta}{d})}{d \Gamma(\frac{2\Delta}{d})} \mathbf{P}_{-1+\frac{\Delta}{d}}(-1+2(rb)^d) \ ,
\end{align}
where $\mathbf{P}$ is a Legendre function of the first kind and $b= d/(4\pi T)$.

To compare more easily with the results in \cite{Yarom:2009mw}, we will work with the bulk viscosity directly. Taking the derivative of $\phi$ with respect to the temperature, evaluating at the horizon, $r=b^{-1}$, and using our formula in terms of the temperature derivative \eqref{temp}, we find
\begin{align}
\zeta = \frac{s}{\pi} v_s^4 \left(\frac{d \Lambda}{4\pi T}\right)^{2(d-\Delta)} \frac{(\Delta-d)^2 \Gamma(\frac{\Delta}{d})^4}{d^2 \Gamma(\frac{2 \Delta}{d})^2} \ .
\end{align}
The entropy density and speed of sound for background are simply (in units where $G^{d+1}_N = (16\pi)^{-1}$)
\begin{align}
s = 4\pi b^{1-d}; \quad v_s^2 = \frac{1}{(d-1)^2} \ .
\end{align}
Using these results and the $\Gamma$ function identity,
\begin{align}
\Gamma(z) \Gamma(z+1/2) = 2^{1-2z} \sqrt{\pi} \Gamma(2z) \ ,
\end{align}
we find
\begin{align}
\left(\frac{d}{4\pi T}\right)^{d-1}\zeta = 16^{1-\Delta/d} \pi \frac{d^{2(d-\Delta-1)}}{(4\pi)^{2(d-\Delta)}} \frac{(d-\Delta)^2}{(d-1)^2} \left(\frac{\Gamma(\frac{\Delta}{d})}{\Gamma(\Delta/d+1/2)}\right)^2 \left(\frac{\Lambda}{T}\right)^{2(d-\Delta)}
\end{align}
which matches Eq. 4 of \cite{Yarom:2009mw}.

\section{Holographic QCD models}

In this section we will consider holographic phenomenological models of QCD, based on gravitational backgrounds with a scalar.
In this setup one starts with one scalar field coupled to gravity in five dimensions with the action
\begin{align}
S = \frac{1}{16\pi} \int d^5 x \sqrt{-g} \left({\cal R} - \frac{1}{2} (\partial \Phi)^2 - V(\Phi) \right) \ . \label{QCDmetric}
\end{align}
One considers the scalar field $\Phi = \phi(r)$, with the metric ansatz
\begin{align}
ds^2 = e^{2A(r)}[-h(r) dt^2 + d \vec{x}^2] + e^{2B(r)} \frac{dr^2}{h(r)} \ .
\end{align}
Following \cite{Gubser:2008ny} one makes the gauge choice that the scalar field itself is the radial coordinate, i.e. $\Phi=r$.

With the choice of metric ansatz, one finds the entropy density
\begin{align}
s = \frac{e^{3A(\Phi_H)}}{4} \ ,
\end{align}
which implies that
\begin{align}
s\frac{d\phi^H}{ds}= \frac{1}{3 A'(\Phi^H)} \ ,
\end{align}
and the formula \eqref{ratio} gives
\begin{align}
\frac{\zeta}{\eta} = \frac{1}{9} A'(\Phi^H)^{-2} \ .
\end{align}

We can go a step further by using the $rr$ metric and the scalar field equation
\begin{eqnarray}
6 A' h' + h(24 A'^2-1) + 2 e^{2B} V = 0, \\
4 A' - B' + \frac{h'}{h} + \frac{e^{2B}}{h} V'  =0 \ ,
\end{eqnarray}
evaluated at the horizon ($h=0$). These imply that $A'(\Phi_H) = V(\Phi_H)/3V'(\Phi_H)$, and thus
\begin{eqnarray}
\frac{\zeta}{\eta} = \left(\frac{V'(\Phi_H)}{V(\Phi_H)}\right)^2. \label{RayQCD}
\end{eqnarray}

This result agrees with the ratio found in \cite{Gubser:2008sz,Kiritsis} using the holographic Kubo formula
\begin{align}
\frac{\zeta}{\eta} = \left(\frac{V'(\Phi_H)}{V(\Phi_H)}\right)^2 |c_b|^2, \label{zetaKubo}
\end{align}
only when the overall factor $c_b =1$. To explain the origin of this factor, we briefly review the analysis of \cite{Gubser:2008sz,Kiritsis}, which starts by considering the Kubo formula
\begin{align}
\zeta = -\frac{1}{9} \lim_{\omega \rightarrow 0} \frac{1}{\omega} lm G^{R}(\omega,0) \ , \label{Kubo}
\end{align}
where $G^R$ is the Fourier transform of the retarded Green's function of the trace of the gauge theory's stress tensor
\begin{align}
G^{R}(\omega, \vec{k}) = -i \int d^3 x dt e^{i \omega t - i \vec{k}\cdot \vec{x}} \theta(t)  <[T_i{}^i(t,\vec{x}),T_j{}^j(0,0)]> \ .
\end{align}
In the AdS/CFT prescription, two point functions of stress tensor operator $T_{\mu \nu}$ can be calculated using the metric perturbations $H_{\mu \nu}$. Using $SO(3)$ invariance and the gauge $\Phi=r$, one can show that the relevant metric perturbation dual to $T_i{}^i$ is
\begin{align}
\delta g = diag(g_{tt},g_{11},g_{11},g_{11},g_{rr}) \ ,
\end{align}
where, to linear order,
\begin{align}
g_{tt} = -h e^{2A}(1+ H_{tt}); \quad g_{11} = e^{2A}(1+H_{11}); \quad g_{rr} = \frac{e^{2B}}{h}(1+ H_{rr}) \ .
\end{align}
Using the ansatz of harmonic time dependence for the perturbations, $e^{-i\omega t} H_{\mu \nu}(\Phi)$, one can show the equation for $H_{11}$ decouples
\begin{align}
H''_{11} = \left(-\frac{1}{3A'}-4A'+3B'+\frac{h'}{h}\right)H'_{11} + \left(-\frac{e^{2B-2A}}{h^2}\omega^2 + \frac{h'}{6hA'}- \frac{h' B'}{h}\right) H_{11} \ . \label{heqn}
\end{align}
To find the bulk viscosity it is sufficient to solve for $H_{11}$. This is done subject to the boundary conditions that $H_{11} \rightarrow 1$ at the boundary and that it is purely infalling at the horizon ($\Phi \rightarrow \Phi_H$):
\begin{align}
H_{11} \rightarrow c_b (\Phi_H - \Phi)^{-\frac{i\omega}{4\pi T}}, \label{horizonBC}
\end{align}
where $c_b$ is some normalization factor. With the solution in hand, one can show that the Kubo formula reduces to just \eqref{zetaKubo}.

Determining the bulk viscosity via the Kubo formula in this class of models amounts to determining $c_b$ by solving numerically \eqref{heqn} with $\omega=0$ and the boundary conditions imposed at the horizon and at infinity. As was shown explicitly in \cite{Gubser:2008sz,Kiritsis}, $c_b$ is not 1 in general and therefore there is a discrepancy between the Kubo formula answer for the bulk viscosity and the value computed with the focusing equation. A plot of $c_b$ as a function of $\lambda_h = e^{\Phi_H}$ is presented in Figure 5 of \cite{Kiritsis}, which shows that the focusing equation agrees with the Kubo formula only in certain limits. One feature of this plot is that $c_b$ approaches unity in the high temperature regime. In this case the horizon is essentially near the boundary.

Another important approximation where the horizon does capture the bulk viscosity is in the \textit{adiabatic} regime discussed in \cite{Gubser:2008ny}. Note that
when the potential is an exponential function of the scalar field (see, for example the Dp branes in Section III) the speed of sound is a constant related to $V'(\Phi_H)/V(\Phi_H)$. In the adiabatic regime, one allows $V'/V$ and the speed of sound to be a slowly varying functions of the horizon value of the scalar field. In terms of the entropy density and temperature, this translates into
\begin{eqnarray}
ln s = - \int^{\Phi_H} d\Phi  \frac{V}{V'} + \cdots \nonumber \\
ln T =  \int^{\Phi_H}  d \Phi \left( \frac{1}{2} \frac{V'}{V} - \frac{1}{3} \frac{V}{V'} \right) + \cdots \label{adiabaticTs}
\end{eqnarray}
where the $\cdots$ represent the slowly varying terms in $\Phi_H$.  In \cite{Kiritsis} it was demonstrated that the fluctuation equation \eqref{heqn} for $H_{11}$ simplifies in adiabatic regime. As a result, the solution for $H_{11}$ is independent of $\Phi$ and the boundary conditions require $c_b = 1$.

The trivial radial behavior of the $\omega=0$ linearized fluctuation equation in the adiabatic regime is reminiscent
 of the behavior of the equation discovered by Iqbal and Liu \cite{Iqbal:2008by} that governs the radial flow of a shear viscosity defined on timelike surfaces of constant $r$ between the horizon and boundary. In the low frequency limit, the radial dependence in this equation always drops out and it is convenient to evaluate the shear viscosity of the boundary gauge theory on the horizon. Our results suggest that a similar equation for the bulk viscosity would only allow one to work at the horizon if the additional condition that the potential and thermodynamic variables are slowly varying functions is imposed.

\section{Discussion}
 \label{sec:discussion}

Using the null focusing equation, we derived a simple novel holographic formula (\ref{ratio}) for the bulk viscosity of
field theory plasmas at strong coupling. The formula was expressed in terms of the dependence of scalar fields
at the horizon  on thermodynamic variables such as the entropy and charge densities. We applied the formula to three classes of gauge theory plasmas: non-conformal branes, perturbations of ${\cal N}=4$ SYM and holographic models of QCD. In the first two classes of models the formula agreed with previous calculations based on quasi-normal modes analysis (or equivalently the holographic Kubo formula) and the hydrodynamic expansion. Note, however, that in all these cases, the calculations of the bulk viscosity in the literature
where performed essentially in the high temperature approximation. In the last class of phenomenological holographic models of QCD, we found an agreement in two approximations: the high temperature limit and the adiabatic approximation.

Let us discuss the range of applicability of the formula.
The null focusing equation encodes dynamics of the horizon. For the boundary field theory this is the IR regime.
The bulk viscosity has in general a non-trivial dependence on the energy scale. For instance, in QCD type theories, dimensional  transmutation
expresses this non-trivial dependence by relating the running coupling $g$ at the temperature scale $T$ to the dynamical scale $\Lambda$
\begin{equation}
\left(\frac{\Lambda}{T}\right)^b = e^{-\frac{1}{g^2(T)}} \ ,
\end{equation}
where $b$ is the one-loop $\beta$-function coefficient.
In such theories, the IR horizon calculation is not expected to capture this running.

In the high temperature limit, the horizon is basically near the boundary and essentially captures also the UV physics; this is the reason for the agreement
between the different types of calculations of the bulk viscosity.
In the adiabatic limit on the other hand, when considering the low frequency limit, the radial dependence in the calculation of the bulk viscosity
drops out and  horizon and boundary calculations yield the same result.
It would be interesting to work in the membrane paradigm formalism to see if this is indeed the case.

A bound on the bulk to shear viscosity ratio has been proposed in \cite{Buchel:2007mf}:
\begin{align}
\frac{\zeta/\eta}{\frac{1}{d-1}-v_s^2} \geq 2 \ .
\end{align}
Interestingly, the bound is satisfied by the formula (\ref{ratio}) in all the cases that we studied in the paper.
In particular, the bound is satisfied in the high temperature expansion used to study the various perturbations of the $\mathcal{N}=4$ SYM, while
in the adiabatic regime,  the formula (\ref{ratio}) saturates the bound. This follows from \eqref{zetaKubo} with $c_b=1$,  together with the approximation \eqref{adiabaticTs} for $v_s^2$.

\section*{Acknowledgements}

The work of Y.O. is supported in part by the Israeli
Science Foundation center of excellence, by the Deutsch-Israelische
Projektkooperation (DIP), by the US-Israel Binational Science
Foundation (BSF), and by the German-Israeli Foundation (GIF).

\section*{Appendix}

Here we show explicitly that the choice of conformal frame in Section III C does not effect the bulk viscosity. The equations of motion for the dual frame action \eqref{ksaction} imply that
\begin{align}
R_{\mu \nu} \ell^\mu \ell^\nu = \frac{1}{2\sigma-d}(D\phi)^2 + \ell^\mu \ell^\nu \nabla_\mu \nabla_\nu \phi \ ,
\end{align}
so that the Raychaudhuri equation has the form
\begin{align}
-\ell^\mu\del_\mu\theta + \kappa\theta - \frac{1}{d-1}\theta^2- \pi_{\mu \nu} \pi^{\mu \nu} =
\frac{1}{2\sigma-d} (D\phi)^2 + \ell^\mu \ell^\nu \nabla_\mu \nabla_\nu \phi .
\label{modfocusing}
\end{align}
Note that $\theta$ is expressed in terms of the area current as before, $v^{-1} \partial_\mu (v \ell^\mu)$. Here, instead we want to work with the entropy current $s^\mu = (1/4) e^\phi v \ell^\mu$. One can define a modified expansion
\begin{align}
\tilde{\theta} = (e^{\phi} v)^{-1} \partial_\mu (e^{\phi} v \ell^\mu) = D \phi + \theta \ .
\end{align}
The entropy current $\partial_\mu s^\mu = s \tilde{\theta}$. Now let's consider (\ref{modfocusing}) at lowest order. First, we rewrite the 2nd \textit{covariant} derivative of $\phi$ in the following way:
\begin{align}
\ell^\mu \ell^\nu \nabla_\mu \nabla_\nu \phi =
\ell^\mu \nabla_\mu(\ell^\nu \nabla_\nu \phi) -
\kappa D \phi .
\end{align}
The first term is now $O(\epsilon^2)$, but the second is first order, by virtue of the geodesic equation $\ell^\mu \nabla_\mu \ell^\nu = \kappa \ell^\nu$. Thus we are left at $O(\epsilon)$ with
\begin{align}
\kappa(\theta+D\phi) = \kappa \tilde{\theta} = 0 .
\label{Jordanideal}
\end{align}
So, as expected at ideal order, the Raychaudhuri equation is equivalent to the vanishing of the entropy current $\partial_\mu (s \ell^\mu)$.

Now let's consider the equation to $O(\epsilon^2)$. We write it in the following form
\begin{align}
2\pi T \tilde{\theta} = D\theta + \frac{1}{d-1} \theta^2 + \pi_{\mu \nu} \pi^{\mu \nu} + \frac{1}{2\sigma-d} (D\phi)^2 + D(D(\phi)) \ .
\end{align}
Now one can impose the ideal equation $\theta = -D\phi$ in the $O(\epsilon^2)$ terms on the right hand side. Expressing the $D\phi$ term again in terms of derivatives of the thermodynamic variables, we find
\begin{align}
\partial_\mu (s \ell^\mu) = \frac{s^3}{2\pi T}~ \frac{2\sigma-1}{(d-1)(2\sigma-d)} \left(\frac{d\phi_H}{ds}\right)^2 + \frac{s}{2\pi T} \pi_{\mu \nu} \pi^{\mu \nu}  \ .
\end{align}
Using the scalar field solution,
\begin{align}
\phi^{H} = (2\sigma-d) \ln (4s)^{\frac{1}{2\sigma-1}} \ ,
\end{align}
one can read off a bulk to shear ratio that agrees with \eqref{NC}.

\end{document}